\documentclass[conference]{IEEEtran}
\IEEEoverridecommandlockouts
\usepackage{cite}
\usepackage{amsmath,amssymb,amsfonts}
\usepackage{algorithmic}
\usepackage{graphicx}
\usepackage{textcomp}
\usepackage{xcolor}
\usepackage{xurl}
\usepackage{tabularx}
\usepackage{array}
\usepackage[hidelinks]{hyperref}
\usepackage{multirow}
\usepackage{multicol}
\def\BibTeX{{\rm B\kern-.05em{\sc i\kern-.025em b}\kern-.08em
    T\kern-.1667em\lower.7ex\hbox{E}\kern-.125emX}}
\begin{document}

\title{A Structured Cyber Threat Intelligence Dataset Using STIX 2.1 Entities and MITRE ATT\&CK Mappings\\
\author{
Dipshikha Das, Arnab Banik, Md. Shariful Islam, and Md Rayhanur Rahman%
\thanks{Dipshikha Das, Arnab Banik, and Md. Shariful Islam are with the
Institute of Information Technology, University of Dhaka, Dhaka, Bangladesh.
E-mail: \{bsse1218, bsse1230, shariful\}@iit.du.ac.bd.}%
\thanks{Md Rayhanur Rahman is with the Department of Computer Science,
The University of Alabama, Tuscaloosa, AL 35487, USA.
E-mail: mrahman87@ua.edu.}
}
}


\maketitle

\begin{abstract}

Cyber threat intelligence (CTI) reports are typically written in unstructured formats, which complicates the extraction and analysis of important entities and adversarial behaviors. Although existing CTI research provides extraction tools, knowledge-graph frameworks, and MITRE ATT\&CK-mapped datasets, curated report-level datasets that preserve complex entity relationships and normalized adversarial behaviors remain limited. To address this limitation, this study presents a manually constructed dataset of 150 English-language CTI reports each represented as  STIX 2.1 based graphs which includes 4,777 STIX entities, 5,817 STIX relationships in total, and 1,273 STIX attack-pattern entities (adversarial-behaviors) mapped to 269 unique MITRE ATT\&CK Enterprise techniques and sub-techniques. Twenty five randomly sampled reports were independently assessed by two cybersecurity researchers, which shows \textit{substantial} inter-rater agreement. Disagreements were subsequently adjudicated to establish a gold-standard reference dataset. Four locally deployed open-source LLMs were evaluated as automated judges against this adjudicated reference sample. Qwen3.6:27B achieved the strongest overall performance, with a maximum kappa score of 0.803, micro-F1 scores exceeding 92\%, and false-positive rates below 5\%. The dataset provides a benchmark for CTI information extraction, knowledge-graph construction, incident analysis, and threat attribution. 
The findings further indicate that locally deployed LLMs can support human reviewers in identifying annotation inconsistencies but expert validation remains essential.
\end{abstract}


\begin{figure*}[!t]
    \centering
    \includegraphics[width=\textwidth]{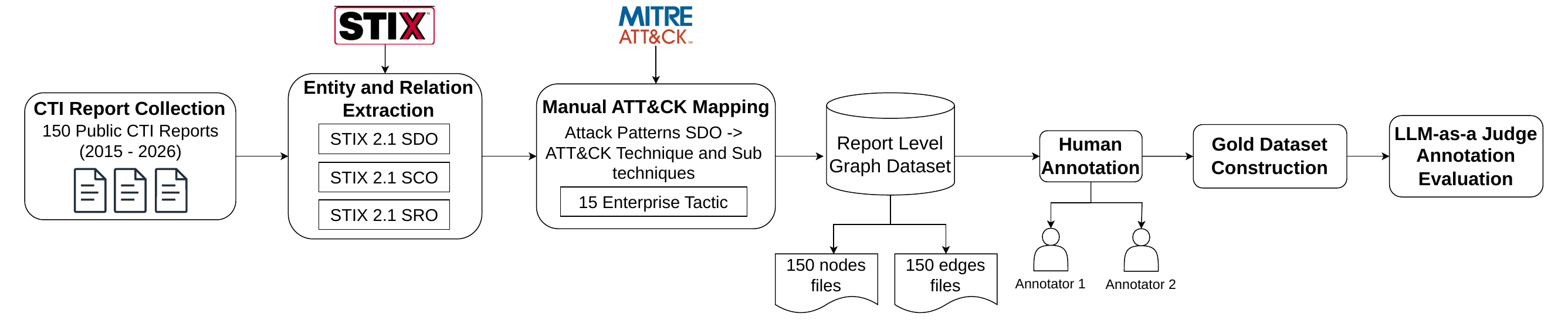}
    \caption{Overall Dataset Construction and Annotation Workflow}
    \label{fig:architecture}
\end{figure*}

\section{Introduction}
Cyber threat intelligence (CTI) reports provide critical information about attackers, adversarial behaviors and malicious operations \cite{cti}. The reports are published by different vendors and are generally written in natural language in unstructured formats. Differences in terminology, writing style, structure, and formatting makes it difficult to analyze and extract actionable intelligence easily  \cite{attackg}. Structured representations can address this problem by organizing CTI information in a standardized format. 
Structured Threat Information Expression (STIX) provides a standardized vocabulary for representing threat entities, cyber observables, and semantic relationships \cite{stix_intro}. On the other hand, MITRE ATT\&CK is a widely used knowledge base that organizes adversarial behaviors into TTPs - Tactics (attackers' intentions), Techniques (methods of attack), sub-techniques (highly specific methods of attack) and lastly Procedure (Specific or step by step implementation) \cite{mitre_attack}. Combining these frameworks provides a consistent format for representing CTI information, which makes information extraction easier, and analysis time shorter.

Previous studies have developed methods for extracting structured threat information from CTI reports. These methods individually identify attack actions or entities and relationships, or MITRE ATT\&CK techniques. Some approaches extract STIX objects and their relationships \cite{stixnet, automated}, while others construct or enrich cybersecurity knowledge graphs \cite{attackg, threatkg, tinker, ctinexus}. Many existing studies also primarily provide automated extraction tools, custom graph representations, text-span annotations, or statement-level ATT\&CK labels \cite{semantic, extractor, ladder, ctihal, annoctr, tram, ttphunter, ttpxhunter}. These representations and tools may not preserve the broader relationships among the entities. Another challenge is that different vendors use their own terminologies for referring to the same adversarial behaviors. As a result, there remains limited availability of curated datasets that represent each CTI report individually through standardized entities, relationships between these entities, and representing adversarial behavior using a normalized format.

This work addresses these gaps by introducing a manually constructed and reviewed dataset of 150 English-language CTI reports, each represented as a directed, heterogeneous graph containing STIX 2.1 Domain Objects (SDO), Cyber-observables (SCO), and STIX Relationship Objects (SRO). SDOs represent a cyber threat concept or entity, SCOs represent observable technical data \cite{stix_intro}. These SDOs and SCOs are represented as nodes. SROs define the relationships among these STIX objects/entities \cite{stix_intro}, and these are represented as edges in the graph structure. The dataset contains 4,777 nodes and 5,817 relationship edges. Among the SDOs, attack-pattern refers to the adversarial behaviors that attackers use to achieve their goal. To address the earlier mentioned problem of differing terminologies for the same attack behavior, each attack-pattern node is manually mapped to MITRE ATT\&CK enterprise techniques and sub-techniques under specific tactics. This produces 1,273 mappings covering 269 unique ATT\&CK identifiers across all 15 Enterprise ATT\&CK tactics. Existing datasets present entities, relationships, or technique labels separately, but this proposed dataset shows how these entities are connected and represents the adversarial behaviors described in each report in a consistent format.

Two cybersecurity researchers independently annotated a randomly sampled subset of 25 reports. A gold annotation set was constructed by resolving the disagreements between the human annotators. Four local LLMs annotated the sample dataset and the results of these LLM's were compared against the gold annotation reference set.     

The main contributions of this work include:
\begin{itemize}
    \item A curated dataset that represents CTI reports as STIX entities and relationships.
    
    \item Manual mapping of adversarial behaviors (STIX attack pattern SDO) to MITRE ATT\&CK Enterprise techniques and sub-techniques.

    \item Human annotation of the sample dataset with \textit{substantial} inter-rater agreement between the annotators \cite{landis1977measurement}, achieving quadratic weighted Cohen's $\kappa_w$ scores of 0.64, 0.67 and 0.62 for Entity Correctness, ATT\&CK Mapping, and Relationship Correctness respectively. 
    
    \item A systematic assessment of local open-source LLMs for automated dataset annotation by following the LLM-as-a-judge \cite{llm_judge} paradigm.
\end{itemize}

\section{Related Work}

Previous CTI research has mainly focused on extracting threat information, constructing knowledge graphs, and mapping attack descriptions to MITRE ATT\&CK TTPs.

Several studies have worked on cybersecurity knowledge graph construction. AttacKG \cite{attackg} constructs attack-behavior graphs and identifies associated ATT\&CK techniques. Its manual evaluation included only 16 reports, which may not fully represent different CTI writing styles and threat domains. The manually reviewed dataset proposed in this work can support broader evaluation of similar extraction and mapping approaches. ThreatKG \cite{threatkg} and TINKER \cite{tinker} extract and integrate threat information from multiple sources to construct cybersecurity knowledge graphs. CTINEXUS \cite{ctinexus} applies in-context learning to construct knowledge graphs from recent CTI reports and can adapt its output to the STIX framework. These systems focus on automated extraction and knowledge aggregation rather than producing a manually reviewed STIX and ATT\&CK-based representation of CTI reports.

Other studies have focused specifically on STIX-based information extraction. STIXnet \cite{stixnet} extracts STIX entities and relationships from unstructured CTI reports. The semantic-chunking approach in \cite{semantic} employs consensus filtering to extract CTI triples, whereas the validation-and-repair framework in \cite{automated} identifies and repairs invalid STIX 2.1 triples. Researchers also examined the extraction and annotation of ATT\&CK TTPs. EXTRACTOR \cite{extractor} extracts attack behaviors, while LADDER \cite{ladder} extracts attack patterns from unstructured reports and maps them to the ATT\&CK framework. CTI-HAL \cite{ctihal} provides human-annotated ATT\&CK mappings for 81 CTI reports and reports substantial inter-annotator agreement \cite{landis1977measurement}. It mainly focuses on statement-level TTP annotations rather than complete STIX-based report representations. AnnoCTR \cite{annoctr} comprises of 400 CTI reports annotated with general entities, including a subset of 120 reports additionally annotated with cybersecurity entities and ATT\&CK concepts. Its annotations similarly focus on individual entities and concepts rather than connected STIX entities and relationships. TRAM \cite{tram}, TTPHunter \cite{ttphunter}, and TTPXHunter \cite{ttpxhunter} identify ATT\&CK techniques from report text but do not extract the entities and relationships needed to capture the full report context.


By combining a graph-structured STIX 2.1 representation with standardized ATT\&CK mappings, this work provides a foundational ground-truth dataset that preserves entity relationships, adversarial behaviors, and the broader contextual information described in modern CTI reports.


\section{Methodology and Dataset Construction}
Figure ~\ref{fig:architecture} illustrates the overall dataset construction workflow, with human annotation and LLM annotation evaluation.


\subsection{Data Collection}
The dataset consists of English-language cyber threat intelligence reports collected mainly from references available in the ATT\&CK knowledge base \cite{mitre_attack}. Additional reports were gathered from public sources, including The DFIR Report \cite{dfir_report}, BleepingComputer \cite{bleepingcomputer}, and other CTI publishing sites. Only reports containing sufficient contextual and technical information were included in the dataset.

\subsection{Entity and Relationship Extraction}
Each report is converted into a structured STIX 2.1 graph representation \cite{stix_intro}. The dataset contains 13 SDOs and 16 SCOs. The extracted SDOs include attack patterns, campaigns, courses of action, identities, indicators, infrastructures, intrusion sets, locations, malware, observed data, threat actors, tools, and vulnerabilities. The extracted SCOs include artifacts, autonomous systems, directories, domain names, email addresses, email messages, files, IPv4 addresses, MAC addresses, network traffic, processes, software, URLs, user accounts, Windows Registry keys, and X.509 certificates.
Relationships between entities are represented using SROs. The dataset supports SDO–SDO, SDO–SCO, and SCO–SDO relationships and follows standard STIX 2.1-defined SROs \cite{stix_intro}. Entities and relationships were extracted only when the source report explicitly supported them. 

\subsection{Mapping Attack Patterns to MITRE ATT\&CK Techniques}
Adversarial behaviors described in natural language were identified as attack patterns SDO in the dataset. These attack patterns were manually mapped to ATT\&CK version 19.1 enterprise techniques and sub-techniques within the defined tactics \cite{mitre_attack}. Manual mapping was used to ensure accurate and context-aware ATT\&CK technique and sub-technique assignments.
As the same adversarial behaviors are described differently across CTI reports,  comparison and analysis of these behaviors would be difficult without proper ATT\&CK mappings. For each behavior, first the corresponding ATT\&CK tactic was identified, followed by identifying the most specific and report-supported technique or sub-technique.

Attack behavior was mapped to most suitable ATT\&CK sub-technique when sufficient evidence was available. Otherwise, it was mapped to the corresponding parent technique.

The ATT\&CK Enterprise matrix comprises of 15 tactics, 222 techniques, and 475 sub-techniques \cite{mitre_attack}. Consequently, manual ATT\&CK mapping was the most challenging task, as it required each observed adversarial behavior to be mapped to the most suitable technique or sub-technique supported by evidence in the source report.

\subsection{Example Data}
Each CTI report consists of two CSV files: a nodes file and an edges file. Each nodes file contains four attributes: a unique report-level identifier (\texttt{id}), the corresponding STIX object category (\texttt{type}), the specific entity name (\texttt{instance}), and contextual details derived directly from the source document (\texttt{description}). The edges file contains \texttt{sourceId}, \texttt{targetId}, and \texttt{relationship}. Each row represents a directed contextual relationship between two entities in the corresponding nodes file. Together, the two files form a STIX-inspired graph representation of each report.

The CTI report in \cite{apt39} describes the cyber-espionage activities of the \textit{Iranian threat actor APT39}.
It initially compromises target systems through spear phishing emails with attachments that deliver a custom variant of \textit{POWBAT malware}. This group also uses \textit{Windows Credential Editor tool} during the attack.

The extracted entities and relationships from this content are shown in Tables~\ref {tab:node-file} and~\ref{tab:edge-file} respectively. 

\begin{figure}[htbp]
    \centering
    \includegraphics[width=0.8\columnwidth]{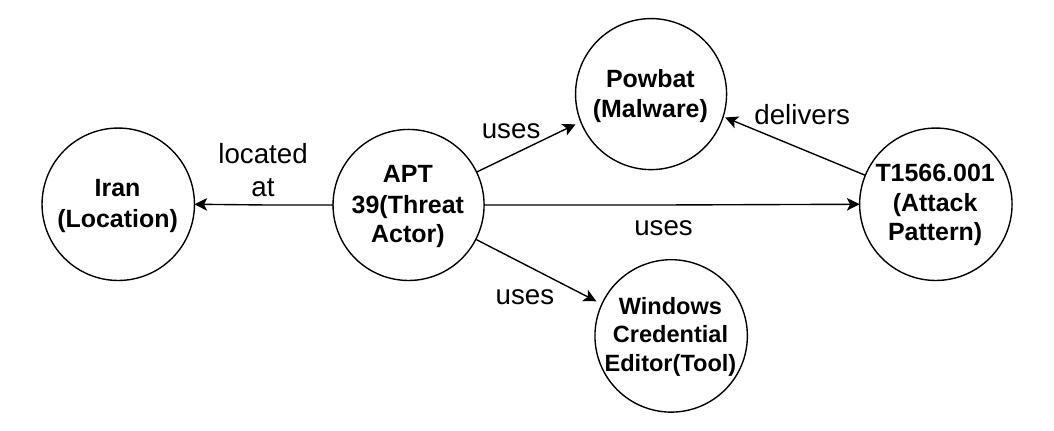}
    \caption{Graph structure of the given example}
    \label{fig:example-graph}
\end{figure}

\begin{table}[htbp]
\caption{Extracted Entities in the Node File}
\label{tab:node-file}
\centering
\scriptsize
\setlength{\tabcolsep}{2.5pt}
\renewcommand{\arraystretch}{1.1}

\begin{tabularx}{\columnwidth}{
|>{\centering\arraybackslash}p{0.05\columnwidth}|
>{\raggedright\arraybackslash}p{0.13\columnwidth}|
>{\raggedright\arraybackslash}p{0.23\columnwidth}|
>{\raggedright\arraybackslash}X|}
\hline

\textbf{ID} &
\textbf{Type} &
\textbf{Instance} &
\textbf{Description} \\
\hline

01 &
Threat actor &
APT39 &
An Iranian state-sponsored cyber-espionage group associated with Chafer that conducts personal-information theft. \\
\hline

02 &
Location &
Iran &
APT39 is assessed with moderate confidence to originate from Iran and operate in support of Iranian national interests. \\
\hline

03 &
Attack pattern &
Phishing: Spearphishing Attachment (T1566.001) &
Tactic: Initial Access (TA0001). APT39 uses spearphishing emails containing malicious attachments. \\
\hline

04 &
Malware &
Powbat &
A custom backdoor variant used by APT39 to establish an initial foothold. \\
\hline

05 &
Tool &
Windows Credential Editor &
A legitimate credential-management tool abused by APT39 for credential harvesting and privilege escalation. \\
\hline
\end{tabularx}
\end{table}
\begin{table}[htbp]
\caption{Extracted Relationships in the Edge File}
\label{tab:edge-file}
\centering
\footnotesize
\renewcommand{\arraystretch}{1.1}
\setlength{\tabcolsep}{4pt}

\begin{tabularx}{0.80\columnwidth}{
|>{\centering\arraybackslash}p{0.17\columnwidth}|
>{\centering\arraybackslash}p{0.17\columnwidth}|
>{\raggedright\arraybackslash}X|}
\hline
\textbf{Source ID} &
\textbf{Target ID} &
\textbf{Relationship} \\
\hline
01 & 02 & located-at \\
\hline
01 & 03 & uses \\
\hline
03 & 04 & delivers \\
\hline
01 & 04 & uses \\
\hline
01 & 05 & uses \\
\hline
\end{tabularx}
\end{table}
The graph representation of the given example can be seen in Figure~\ref{fig:example-graph}, where instance and type represents the nodes and relationship represents the edges.

\subsection{Human Annotation and Inter-Rater Reliability}
\label{annotation_methodology}


25 reports (approximately 17\% of the dataset) and their corresponding nodes and edges were randomly sampled to establish a human-annotated reference set. The sample set contains 566 nodes, 677 edges, and 147 attack-pattern nodes mapped to Enterprise ATT\&CK techniques and sub-techniques. A graduate level and a post-graduate level cybersecurity researcher independently annotated each node and edge. The annotators followed the rubrics presented in Table~\ref{tab:annotation-criteria}. 
\begin{table}[htbp]
\caption{Human-Annotation Rubrics}
\label{tab:annotation-criteria}
\centering
\footnotesize
\renewcommand{\arraystretch}{1.1}
\begin{tabularx}{\columnwidth}{
|>{\raggedright\arraybackslash}p{0.29\columnwidth}|
>{\raggedright\arraybackslash}X|}
\hline
\textbf{Rubrics} & \textbf{Evaluation Question} \\
\hline
A: Entity type correctness & The instance and its description align with the assigned STIX SDO/SCO types and are supported by the report. \\
\hline
B: Attack-mapping correctness &
The assigned ATT\&CK technique or sub-technique aligns with its official MITRE ATT\&CK definition. \\
\hline
C: Relationship correctness & The relationship between the source and target is supported by the report and aligns with STIX SRO definitions. \\
\hline
\end{tabularx}
\end{table}

\label{annotation_label}
Each node and edge was annotated on a 3-point ordinal scale. The \texttt{supported} label indicates that the existing node/edge is fully consistent with the source report and their definitions. The \texttt{partially\_supported} label indicates that the node/edge that has been extracted from the report is not fully supported by its definitions. The \texttt {unsupported} label indicates that the node/edge is not substantiated by the report according to its definitions.

\label{human-annotation-results}

The quadratic weighted Cohen's Kappa Score  \textbf{$\kappa_w$} is \textbf{0.64} for Type Correctness, \textbf{0.67} for ATT\&CK Mapping Correctness, and \textbf{0.62} for Relationship Correctness. All 3 Rubrics indicate a \textbf{\textit{substantial}} agreement between the annotators \cite{landis1977measurement}. Table \ref{tab:hvh-agreement-matrix} presents the Agreement Matrix where S., P.S. and U. refers to Supported, Partially Supported and Unsupported respectively:


\begin{table}[htbp]
\caption{Human vs Human Agreement Matrix}
\label{tab:hvh-agreement-matrix}
\centering
\footnotesize
\setlength{\tabcolsep}{2pt}
\renewcommand{\arraystretch}{1.08}

\begin{tabular}{l|ccc|ccc|ccc}
\hline
& \multicolumn{3}{c|}{Type} & \multicolumn{3}{c|}{ATT\&CK}  & \multicolumn{3}{c}{Edge}   \\

              & \textit{S.} & \textit{P.S} & \textit{U.} & \textit{S.} & \textit{P.S} & \textit{U.} & \textit{S.} & \textit{P.S} & \textit{U.} \\
\hline
\textit{S.}   & 492         & 6            & 12          & 113         & 3            & 3           & 511         & 41           & 3           \\
\textit{P.S.} & 3           & 24           & 3           & 3           & 13           & 2           & 33          & 13           & 30          \\
\textit{U.}   & 8           & 3            & 15          & 3           & 1            & 6           & 5           & 30           & 11         \\
\hline
\end{tabular}
\end{table}

Disagreements were discussed and resolved through mutual agreement between the annotators. Feedback from the annotators was used to revise and improve the annotated samples. The summary of changes is presented in Table \ref{tab:change-summary}. 

\begin{table}[htbp]
\caption{Sampled Dataset Correction Summary}
\label{tab:change-summary}
\centering
\footnotesize
\setlength{\tabcolsep}{4pt}
\renewcommand{\arraystretch}{1.08}
\begin{tabular}{|l|c|c|c|}
\hline
                               & Type & ATT\&CK & Edge \\ \hline
Old Items Count                      & 566  & 147    & 677  \\ \hline
Total Dropped Items            & 21   & 9      & 71   \\ \hline
Total Updated Items            & 53   & 25     & 95   \\ \hline
\% of Items Changed       & 13.1 & 23.1   & 24.5 \\ \hline
Gold \textit{Supported} Count           & 525  & 132    & 598  \\ \hline
Gold \textit{Partially Supported} Count & 20   & 6      & 8    \\ \hline
Gold \textit{Unsupported} Count         & 0    & 0      & 0    \\ \hline
\end{tabular}
\end{table}



\section{DATASET CHARACTERISTICS}
\subsection{Dataset Scale and Coverage}
The proposed dataset contains 150 English language CTI reports published between 2015 and 2026. It includes 4,777 entities. It also has 5,817 best-suited directed relationships. The summary of the dataset is shown in Table \ref{tab:dataset-summary}.
\begin{table}[htbp]
\caption{Summary of the Proposed CTI Dataset}
\label{tab:dataset-summary}
\centering
\footnotesize
\renewcommand{\arraystretch}{1.1}

\begin{tabularx}{0.9\columnwidth}{
|>{\raggedright\arraybackslash}X|
>{\raggedright\arraybackslash}p{0.22\columnwidth}|}
\hline
\textbf{Characteristic} &
\multicolumn{1}{c|}{\textbf{Value}} \\
\hline
CTI reports & 150 \\
\hline
Publication period & 2015--2026 \\
\hline
Total Valid entities \& relationships & 4,777 \& 5,817\\
\hline
Average nodes \& relationships per Report & 32 \& 39 \\
\hline
Unique ATT\&CK IDs & 269 \\
\hline
Total ATT\&CK occurrences & 1,273 \\
\hline
\end{tabularx}
\end{table}

\subsection{Entity And Relationship Distribution}
The dataset contains 3,446 SDOs and 1,331 SCOs. \textit{Attack-pattern} and \textit{file} are the most frequent SDO and SCO respectively. In contrast, \textit{vulnerability} is the least frequent SDO with 50 occurrences, while \textit{mac-addr} is the least frequent SCO, occurring once. Among the relationships, \textit{uses} occurs the most. The least frequent SRO types are \textit{uploaded-on, derived-from, characterizes, disables}, and \textit{identifies}, each with one occurrence. Table \ref{tab:top-stix-types} presents top 5 most frequent entities and relationships in the dataset.

\begin{table}[htbp]
\caption{Top Five Most Frequent SDO, SCO, and SRO Types}
\label{tab:top-stix-types}
\centering
\footnotesize
\renewcommand{\arraystretch}{1.1}
\setlength{\tabcolsep}{2.5pt}

\begin{tabular}{|l|r|l|r|l|r|}
\hline
\textbf{SDO Type} & \textbf{Freq.} &
\textbf{SCO Type} & \textbf{Freq.} &
\textbf{SRO Type} & \textbf{Freq.} \\
\hline
attack-pattern & 1,272 & file & 550 & uses & 2,697 \\
\hline
identity & 443 & domain-name & 318 & targets & 748 \\
\hline
malware & 354 & ipv4-addr & 172 & indicates & 280 \\
\hline
location & 293 & url & 145 & mitigates & 275 \\
\hline
tool & 230 & software & 47 & based-on & 263 \\
\hline
\end{tabular}
\end{table}

\subsection{MITRE ATT\&CK Coverage}
The dataset contains 269 unique MITRE ATT\&CK technique and sub-technique identifiers, comprising of 123 parent techniques and 146 sub-techniques. In total, the dataset contains 749 parent techniques and 524 sub-techniques. The five most frequently mapped attack patterns are \textit{Ingress Tool Transfer} (T1105), \textit{Data Encrypted for Impact (T1486)}, \textit{Obfuscated Files or Information (T1027)}, \textit{Phishing (T1566)}, and \textit{Exfiltration Over C2 Channel (T1041)}.
The mappings span all 15 ATT\&CK tactics. The tactic \textit{Stealth} contains the highest number of mapped rows and unique techniques, followed by \textit{Initial Access}, \textit{Command and Control}, and \textit{Execution}. 

Overall, the dataset offers broad structural, and behavioral coverage of each CTI report, which can support diverse CTI analysis, information extraction, automated evaluation tasks.

\section{Assessing LLM-as-a Judge for Dataset Annotation}

\subsection{LLM-as-a-Judge Framework}
Independently re-annotating the complete dataset by multiple cybersecurity experts is impractical because of its cost and time consumption\cite{annotation}. Therefore, LLMs are employed as automated annotators to provide a scalable, and complementary solution by following the LLM-as-a-judge paradigm \cite{llm_judge}.

\subsection{Models Used}
4 reasoning models--\texttt{qwen3.6:27b}, \texttt{qwen3:14b}, \texttt{gpt-oss:20b}, and \texttt{gemma4:31b}-- were run locally through Ollama with provider set default parameters. All 4 models were using \texttt{Q4\_K\_M} quantization, and \texttt{4-bit KV-cache} with a maximum prompt size of 32,768 tokens. The models were run on an external \textit{Nvidia RTX 3090} graphics card with 24GB VRam. Local LLMs were selected over commercial APIs to avoid provider-side model changes \cite{llm_drift}, and reduce cost on repeated runs as the dataset expands.

\subsection{Prompting}
The LLMs received the same rubrics defined in section \ref{tab:annotation-criteria} as system prompts, and annotated on the same Ordinal Scale as the human annotators. The LLMs were also provided with official definitions as reference text. Chain-of-Thought(CoT) reasoning \cite{cot} was used for all four models. Each model received batches of items (a maximum of 5 items for \texttt{qwen3:14b and gpt-oss:20b}, and 8 items for \texttt{qwen3.6:27b and gemma4:31b}), and each batch was a standalone LLM call with only the references of the items included in the batch. The output was a \texttt{json array} of the following schema: \texttt{
\{
  reasoning:          string, 
  unsupported\_claims: [string], 
  confidence:         enum(low, medium, high), 
  verdict:            enum(supported, partially\_supported, unsupported)
\}
}


\subsection{LLM Assessment Results}
All four LLMs evaluated the pre and post corrected versions of the 25 samples described in Section \ref{annotation_methodology}. 

Table~\ref{tab:before-results} presents the pre-corrected sample
results. The performance is measured using micro-F1,
macro-precision, macro-recall, macro-F1, and quadratic-weighted
Cohen's Kappa scores against the gold reference set (including \textit{unsupported} class). Micro-F1 measures overall classification
performance across all instances, while the macro metrics give equal importance to all three classes.

\begin{table}[htbp]
\caption{LLM Assessment Results for Pre-Corrected Samples}
\label{tab:before-results}
\centering
\scriptsize
\setlength{\tabcolsep}{2pt}
\renewcommand{\arraystretch}{1.08}
\resizebox{\columnwidth}{!}{
\begin{tabular}{llccccc}
\hline
\textbf{Rubric} & \textbf{Model} &
\textbf{Micro-F1} & \textbf{Macro-P} &
\textbf{Macro-R} & \textbf{Macro-F1} &
\boldmath$\kappa_w$ \\
\hline
\multirow{4}{*}{Type}   & qwen3:14b   & 0.913 & 0.416 & 0.468 & 0.437 & 0.414 \\
                        & gpt-oss:20b & 0.924 & 0.463 & 0.403 & 0.418 & 0.258 \\

                        & qwen3.6:27b & \textbf{0.949} & \textbf{0.611} & \textbf{0.550} & \textbf{0.575} & \textbf{0.675} \\
                        & gemma4:31b  & 0.920 & 0.439 & 0.471 & 0.454 & 0.421 \\
\hline
\multirow{4}{*}{ATT\&CK}    & qwen3:14b   & 0.918 & \textbf{0.729} &0.707 & 0.711 & 0.661 \\
                            & gpt-oss:20b & 0.891 & 0.502 & 0.538 & 0.518 & 0.662 \\

                            & qwen3.6:27b & \textbf{0.939} & 0.709 & \textbf{0.765} & \textbf{0.719} & \textbf{0.803} \\
                            & gemma4:31b  & 0.918 & 0.603 & 0.636 & 0.600 & 0.703 \\
\hline
\multirow{4}{*}{Edge}   & qwen3:14b	    & 0.790	& 0.470	& 0.592	& 0.475 & 0.452 \\
                        & gpt-oss:20b	& 0.809	& 0.477	& 0.583	& \textbf{0.497}	& \textbf{0.467} \\
                        & qwen3.6:27b	& \textbf{0.838}	& \textbf{0.497}	& \textbf{0.597}	& 0.469	& 0.346 \\
                        & gemma4:31b	& 0.799	& 0.427	& 0.523	& 0.436	& 0.248 \\ 

\hline
\end{tabular}}
\end{table}

Table~\ref{tab:after-results} presents the post-corrected sample results. As the post-correction reference
labels do not contain the \textit{unsupported} class, Kappa and macro-averaged metrics are not applicable. Therefore, the performance is reported using micro-F1 and the false-positive rate (FPR) for the \textit{unsupported} class.

\begin{table}[htbp]
\caption{Post-Correction LLM Assessment Results}
\label{tab:after-results}
\centering
\footnotesize
\setlength{\tabcolsep}{4pt}
\renewcommand{\arraystretch}{1.08}
\begin{tabular}{llcc}
\hline
\textbf{Rubric} & \textbf{Model} &
\textbf{Micro-F1} & \textbf{Unsupported FPR} \\
\hline
\multirow{4}{*}{Type}   & qwen3:14b   & 0.941 & 0.029 \\

                        & gpt-oss:20b & 0.934 & 0.031 \\
                        & qwen3.6:27b & \textbf{0.958} & \textbf{0.006} \\
                        & gemma4:31b  & 0.949 & 0.013 \\
\hline
\multirow{4}{*}{ATT\&CK}    & qwen3:14b   & 0.928 & 0.036 \\
                            & gpt-oss:20b & \textbf{0.935} & 0.043 \\
                            & qwen3.6:27b & 0.928 & \textbf{0.029} \\
                            & gemma4:31b  & 0.906 & 0.072 \\
                            
\hline
\multirow{4}{*}{Edge}   & qwen3:14b	    & 0.799 & 0.127  \\
                        & gpt-oss	    & 0.804	& 0.155	 \\
                        & qwen3.6:27b	& \textbf{0.929}	& \textbf{0.031}	\\
                        & gemma4:31b	& 0.880	& 0.089	 \\

\hline
\end{tabular}
\end{table}

\subsection{Analysis of Results}

In the pre-corrected assessment, \texttt{qwen3.6:27b} achieves the highest macro-F1 scores for all 3 rubrics. It also achieves the highest $\kappa_w$ scores of 0.675, and 0.803 on \textit{Type}, and \textit{ATT\&CK} rubrics which indicate \textit{substantial} and \textit{almost perfect} agreement with the gold standard annotation respectively \cite{landis1977measurement}. For \textit{Edges} rubric model performance varies across the evaluation metrics, \texttt{qwen3.6:27b} achieves the highest micro-F1, but \texttt{gpt-oss:20b} achieves the highest macro-F1, and $\kappa_w$ scores.  

All Micro-F1 scores signal that the models' accuracy is high for our single label multiclass gold reference set, and the lower macro average score in comparison properly reflects the class imbalance in the gold set. The lower $\kappa_w$ scores on the \textit{Edges} reflect the reasoning limitations of small to medium sized local LLMs, where the the LLMs try to find verbatim evidence for relationships from a given list of accepted \texttt{sourceId}, \texttt{targetId}, and \texttt{relationship} verb as reference.

In the post-corrected results, \texttt{qwen3.6:27b} again achieves the highest micro-F1 , and the lowest false positive rates for \textit{Type} and \textit{ATT\&CK} rubrics respectively. For \textit{Edges}, \texttt{gpt-oss} scores the highest micro-F1, and \texttt{qwen3.6} scores the lowest false positive rate. All models scored above 93\% in micro-F1, and \texttt{qwen3.6} scored below 5\% in false positive rates. These results show that the model rarely misclassifies the corrected annotations as \texttt{unsupported}.


\section{Discussion and Limitations}
The proposed dataset represents each CTI report in a structured format that represents entities through STIX 2.1 objects and normalizes adversarial behaviors through ATT\&CK mappings. The substantial inter-rater agreement across all three rubrics confirms that the resulting dataset is a reliable and consistent representation of the source reports \cite{landis1977measurement}. The correction rates indicate that ATT\&CK mapping and relationship extraction are more challenging than entity extraction. These tasks require substantial contextual interpretation because adversarial behaviors and relationships are often described implicitly or using vendor-specific terminology. 

The LLM annotation evaluation results reflect similar challenges. High micro-F1 but lower macro-F1 and $\kappa_w$ scores indicate class imbalance and weaker performance on minority labels. \texttt{Qwen3.6:27b} achieved the strongest overall performance, while relationship assessment remained comparatively difficult. So local LLMs may assist human reviewers in identifying inconsistencies but cannot replace expert validation. 
The complete dataset, source code, evaluation prompts and other materials will be made publicly available upon acceptance.

The current dataset presents the following constraints: it is limited to 150 public, English language reports, and it does not model temporal attack sequences described in the CTI reports. Direct human assessment covers approximately 17\% of the dataset. This subset was rigorously corrected to ensure the reference set contains zero unsupported instances.

\section{Conclusion and Future Work}
This study introduced a manually constructed dataset of 150 CTI reports each represented as directed heterogeneous STIX 2.1 graph based structure. The dataset contains 4,777 entities, 5,817 relationships, and 1,273 adversarial behavior (attack pattern SDO) occurrences mapped to 269 unique MITRE ATT\&CK techniques and sub-techniques across all 15 Enterprise tactics. Human assessment of randomly sampled 25 reports indicates \textit{substantial} inter-rater agreement across the 3 annotation rubrics. Evaluation against the adjudicated gold reference set further showed that locally deployed open-source LLMs, particularly \texttt{Qwen3.6}, can support automated scalable annotation, although expert validation remains necessary.

A dataset of this kind with structured STIX representation is valuable for consistently representing complex threat entities, their relationships, and adversarial behaviors described using different terminology across CTI reports. Its standardized and context-preserving representation can support the development and evaluation of CTI report related automated tasks.

Future work may incorporate multilingual and diverse sources, and temporal relationships described in the reports. Additional LLMs and prompting strategies may also be evaluated, together with downstream tasks such as information extraction, incident analysis, CTI report clustering, threat attribution, and knowledge-graph construction.

\end{document}